\documentclass[aps,prl,reprint,showpacs]{revtex4-1}
\usepackage[dvipdfmx]{graphicx}
\usepackage{mediabb}
 
\newcommand{\nn}{\nonumber}

\usepackage{url}

\usepackage{amsmath,amssymb}
\usepackage{color}

\usepackage{bm}

\def\v#1{\mathbf{#1}}			

\def\vv{\v{v}}

\def\vv{\v{v}}

\def\del{\partial}

\usepackage[normalem]{ulem}

\begin{document}


\title{Spin hydrodynamic generation in graphene}


\author{Mamoru Matsuo$^{1,2,3,4}$, Denis A. Bandurin$^{5}$, Yuichi Ohnuma$^{1}$, Yasumasa Tsutsumi$^{3,6}$ and Sadamichi Maekawa$^{3,1}$}
\affiliation{%
${^1}$Kavli Institute for Theoretical Sciences, University of Chinese Academy of Sciences, Beijing, 100190, China.\\
${^2}$CAS Center for Excellence in Topological Quantum Computation, University of Chinese Academy of Sciences, Beijing 100190, China.\\
${^3}$RIKEN Center for Emergent Matter Science (CEMS), Wako, Saitama 351-0198, Japan.\\
${^4}$Advanced Science Research Center, Japan Atomic Energy Agency, Tokai, 319-1195, Japan.\\
${^5}$Moscow Institute of Physics and Technology (State University), Dolgoprudny 141700, Russia.\\
${^6}$Department of Physics, Kwansei Gakuin University, Sanda, Hyogo 669-1337, Japan.
}%

\date{\today}

\begin{abstract}
Graphene hosts an ultra-clean electronic system with electron-electron collisions being the dominant source of scattering above liquid nitrogen temperatures. 
In this regime, the motion of the electron fluid resembles the flow of classical liquids and gases with high viscosity. Here we show that such a viscous electron flow can cause the generation of a spin current perpendicular to the direction of flow. Combining the Navier-Stokes equations and the spin diffusion equation in the presence of the spin-vorticity coupling, we derive an expression for the spin accumulation emerging purely as a result of the viscous electron flow. We explore Poiseuille flow and Jeffery-Hamel flow and show that the spin Hall angle may exceed 0.1 over a wide range of temperatures and can be controlled by carrier density, temperature, and the geometry of sample boundaries. Our theory points to a new functionality of graphene as a spin current source. 

\end{abstract}

\maketitle 

\paragraph{Introduction.---}
Generation, electric control and detection of spin currents are the key tasks in the field of spintronics\cite{Zutic2004,MaekawaEd2012}. 
In magnetic materials, spin current generation can be achieved in a variety of ways: spin accumulation at the interface between ferromagnetic and nonmagnetic materials\cite{Jedema2001}, magnetization dynamics in a ferromagnet excited by a microwave irradiation\cite{Mizukami2002, Tserkovnyak2005}, thermal gradients\cite{SSE1,SSE1a,SSE2}, and phonon dynamics in a ferromagnet\cite{ASP,ASP2,ASP3}.
In a non-magnetic material, the spin current can be generated by the spin Hall effect\cite{Valenzuela2006,Saitoh2006,NLSV}, if the material is characterized by  substantial spin-orbit coupling.

Recently, an alternative method for generating spin currents that exploits the coupling between spin and vorticity was proposed\cite{SAW-SC:Matsuo} and experimentally demostrated in conducting fluids\cite{Takahashi2016}, elastic metals\cite{Kobayashi2017}, and a gradient material\cite{Okano2019}.
The spin-vorticity coupling is responsible for angular momentum conversion between electron spin and mechanical angular momentum of the rotating media. 
This coupling does not rely on the the magnetic moments and spin-orbit coupling and originates from the spin connection in generally covariant Dirac equation based on local Poincar\'{e} invariance, leading to the local angular momentum conservation law of the system\cite{Hehl1976}.
The angular momentum conversion via spin-vorticity coupling expands the choice of materials for spin current generation such as liquid metals\cite{Takahashi2016} as well as materials with a weak spin-orbit coupling like Cu\cite{Kobayashi2017,Okano2019}.
\begin{figure}[hbtp]
	\begin{center}
   \includegraphics[scale=0.4]{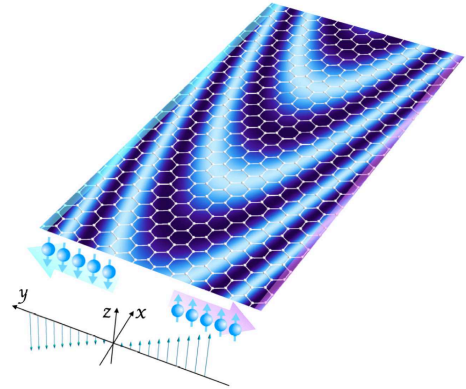}
	\caption{ Spin hydrodynamic generation due to parallel flow in graphene. By applying electric voltage along the $x$-axis, the vorticity gradient is created along the $y$-axis due to the viscous flow of electron fluid. Spin current is generated along the vorticity gradient via the spin-vorticity coupling. As a result, spins accumulate at the graphene edges.  
    }\label{fig:SHA}
	\end{center}
\end{figure}

Graphene, a two-dimensional sheet of carbon atoms, has recently emerged as an exceptional platform for  spintronics\cite{Kawakami2014}. Thanks to weak spin-orbit coupling and vanishing hyperfine interaction, graphene devices have demonstrated ultra-long spin lifetimes reaching tens of nanoseconds\cite{Drogler,Ingla-Aynes2015} and together with recent advances in assembling van der Waals heterostructures enabled the development of spin field-effect switches\cite{Yan2016}. 
Despite concerted efforts, the variety of methods to generate spin-polarized currents in graphene is rather scarce and is usually achieved by two methods: direct spin injection from ferromagnetic contacts or by the spin Hall effect\cite{Kawakami2014}. While the former requires the deposition of air-sensitive materials\cite{Tombros07}, such as cobalt, the latter implies modification of the graphene lattice by point defects  aimed to induce finite spin-orbit coupling (e.g. \cite{Balakrishnan}), which in turn, leads to a reduction of the spin diffusion length. Therefore, the search of alternative mechanisms of spin current generation in graphene remain of high importance.

Simultaneously with the advances in graphene spintronics, 
a seemingly unrelated topic---electron hydrodynamics---emerged\cite{Moll2016,Bandurin2016,Crossno2016}. 
Electron hydrodynamics addresses the behaviour of charged electron fluids, 
in which electron-electron collisions are the dominant scattering source, rendering local thermodynamic equilibrium and ensuring that the behaviour of such systems can be conveniently described by the laws of classical fluid mechanics\cite{gurzhi63}. Thanks to weak electron-phonon coupling, ultra-clean graphene devices offer an exceptional venue to witness hydrodynamic flow of  electron fluid\cite{Lucas2018}. A host of new phenomena such as negative local resistance\cite{Bandurin2016,Bandurin2018,Torre2015,Levitov2016}, superballistic conduction\cite{Guo,Kumar2017}, violation of the Wiedeman-Franz law\cite{Crossno2016}, quantum-critical conductivity\cite{Gallagher2019} and anomalous viscous magnetotransport\cite{Berdyugin2018} have been predicted and unveiled in graphene making electron hydrodynamics a novel paradigm of electron transport in solid state devices.

In this work, we bridge these two seemingly-distant fields: spintronics and electron hydrodynamics and show that such a merge offers an alternative way for the spin current generation in graphene devices. By combing the Navier-Stokes equations of electron hydrodnamics in graphene and the spin diffusion equation in the presence of the spin-vorticity coupling, we obtain the spin accumulation and spin current in graphene in the case of Poiseuille flow and Jeffery-Hamel flow. Our results reveal a new functionality of ultra-clean graphene as a platform for non-magnetic spintronics.

\paragraph{Stokes equation and spin diffusion equation in graphene.---}
As discussed in ~\cite{HydroGraphene,Torre2015,Levitov2016}, at elevated temperatures, electron transport in doped graphene can be described by the hydrodynamic equations:
\begin{eqnarray}
\nabla \textbf{J}(\textbf{r})=0, \,\,
-\frac{n e}{m} \nabla \Phi + \nu \nabla^2 \v{v} =  0. \label{eq:NS}
\end{eqnarray}
Here $\textbf{J}(\textbf{r})$ and $\Phi(\textbf{r})$ are the linearised particle current density and the electric potential in the 2D graphene plane respectively, $e$ and $m$ are the electron charge and the effective mass and $\nu$ is the kinematic viscosity of graphene's electron fluid. Note, the only pseudo-relativistic correction that remains in Eq.~(\ref{eq:NS}) is $m= k_{\rm F}/v_{\rm F}$ expressed through the Fermi wave number $k_{\rm F}$ and the Fermi velocity $v_{\rm F}$  ~\cite{Torre2015}. Equation~(\ref{eq:NS}) states the balance of viscous friction and electric force $eE=-e\nabla\Phi$ and resembles the familiar Stokes equation for classical fluids. 

The spin-diffusion equation in the presence of a viscous flow is derived by using the quantum kinetic theory in Ref.~\cite{SHD:Matsuo} as 
\begin{eqnarray}
(\del_t -D_{\rm s} \nabla^2 + \tilde\tau_{\rm sf}^{-1}) \delta \mu_S =  -\hbar \tilde\tau_{\rm sf}^{-1} \zeta \omega,\label{g.spin-diff}
\end{eqnarray}
where $\delta \mu_S$ is the spin accumulation, $D_{\rm s}$ is the diffusion constant, $\tilde\tau_{\rm sf}$ is the spin relaxation time, and $\zeta$ is the renormalization factor of the spin-vorticity coupling~\cite{SHD:Matsuo}.
The spin-vorticity coupling arises in the Dirac Hamiltonian in the local rest frame of the fluid without the non-relativistic limit~\cite{SHD:Matsuo}. We note that the spin-orbit coupling is derived from the non-relativistic limit. As a result, the magnitude of the bare spin-vorticity coupling is the order of $\mathcal{O}(1)$ whereas the bare spin-orbit coupling is $\mathcal{O}(1/c^2)$ with the speed of light $c$. 
In addition, the spin-vorticity coupling is enhanced by band structure effects as the spin-orbit coupling~\cite{RenormSRC}.  
This implies that the spin dynamics driven by the spin-vorticity coupling is potentially larger than that driven by the spin-orbit coupling~\cite{Doornenbal2019}. Previously, in order to probe spin-vorticity coupling, the vorticity fields were created by mechanical excitation of the fluid motion~\cite{Takahashi2016} or by elastic deformations~\cite{Kobayashi2017} under the assumption that electrons adiabatically follow the motion or deformations.
Recently, the spin current generation due to the transfer of angular momentum from the vorticity of electron flow in surface-oxidized Cu is experimentally demonstrated\cite{Okano2019}.
In contrast, in doped graphene, the vorticity field emerges naturally as a result of viscous electron flow~\cite{Levitov2016,Torre2015}. 
In the case of both the surface-oxidized Cu and the graphene in the presence of the viscous electron flow, the diffusive transport of spins is affected by the vorticity field of the non-uniform electron flow. In this situation, the non-uniform vorticity field couples to the conduction electron spins to assure the local angular momentum conservation law, originating from the local rotational symmetry\cite{Hehl1976}. Therefore, we should include the spin-vorticity coupling into the spin diffusion equation as discussed in previous studies\cite{SHD:Matsuo,Okano2019}.

Here, we focus on hydrodynamics of the electron fluid in graphene in the continuum limit. Gauge fields in graphene in the continuum model has been studied in Ref.~\cite{Vozmediano} which also discusses spin connection. The flavors, i.e.~the valley and pseudo spin degrees of freedom, in graphene are included in the Dirac Hamiltonian~\cite{Vozmediano}. In this paper, the real spin is also considered.

\paragraph{Renormalization factor for doped graphene.---}
The renormalization factor of the spin-vorticity coupling in Eq.~(3) is given by~\cite{SHD:Matsuo}
\begin{align}
    \zeta=\frac{\int d^2k\omega_{\bm k}({\bm r},t)\tilde\tau_{\rm sf}({\bm k})^{-1}\partial_k f_k^0}{\omega({\bm r},t)\tilde\tau_{\rm sf}( k_{\rm F})^{-1}\int d^2k\partial_k f_k^0},
    \label{zeta}
\end{align}
which is obtained from the linear response of conduction electron spins to the fluid vorticity~\cite{Takahashi2016}. An advantage of the electron fluid in graphene is that the renormalization factor can be calculated from microscopic descriptions.
For doped graphene, since the group velocity of a quasiparticle with energy $\epsilon_{\bm k}=\hbar{\bm v}_{\rm F}\cdot{\bm k}$ is the Fermi velocity ${\bm v}_{\rm F}$,
the contribution of the fluid velocity by each quasiparticle is ${\bm v}_{\rm F}f_{\bm k}({\bm r},t)$ where $f_{\bm k}({\bm r},t)$ is the quasiparticle distribution function.
The vorticity $\omega({\bm r},t)$ is obtained by summing up contributions from each of the quasiparticles in a Brillouin zone:
    $\omega({\bm r},t)
    =\sum_{\bm k}\omega_{\bm k}({\bm r},t) =\frac{1}{4k_{\rm F}^2}\int d^2k\omega_{\bm k}({\bm r},t) \nn
    = \frac{1}{4k_{\rm F}^2}\int d^2k({\bm\nabla}\times{\bm v}_{\rm F}\delta f_{\bm k}({\bm r},t))$,
where $\delta f_{\bm k}({\bm r},t)\equiv f_{\bm k}({\bm r},t)-f_k^0$ is the deviation of the distribution from the Fermi distribution $f_k^0$ by an external electric potential $\Phi$.
Thus, the vorticity is created by the quasiparticles excited by the electric potential.

Next, we assume that the spin-flip relaxation time $\tilde\tau_{\rm sf}({\bm k})$ is isotropic for all direction of ${\bm k}$ and less sensitive to the wave number around the Fermi wave number $k_{\rm F}$.
When a small electric potential is applied, that is, quasiparticles are excited in the vicinity of the Fermi level, the renormalization factor in \eqref{zeta} is independent of the spin-flip relaxation time:
 $   \zeta=\frac{4k_{\rm F}^2}{\int d^2k\partial_k f_k^0}\frac{\int d^2k\omega_{\bm k}({\bm r},t)\partial_k f_k^0}{\int d^2k\omega_{\bm k}({\bm r},t)}
$.
 For a small electric potential $|\Phi|\ll k_{\rm B}T$, the derivative of the Fermi distribution in the numerator can be approximate to the value at the Fermi wave number:
 $   \int d^2k\omega_{\bm k}({\bm r},t)\partial_k f_k^0\approx\left(\partial_k f_k^0\right)_{k=k_{\rm F}}\int d^2k\omega_{\bm k}({\bm r},t)$.
Therefore, the renormalization factor reads
\begin{align}
    \zeta=\frac{1}{2\pi}\frac{T_{\rm F}}{T},
\end{align}
where $T_{\rm F}=\hbar v_{\rm F}k_{\rm F}/k_{\rm B}$ is the Fermi temperature.

In the following, the spin hydrodynamic generation in graphene will be shown by combining Eqs. (\ref{eq:NS}) and (\ref{g.spin-diff}).

\paragraph{Spin Hall angle for Poiseuille flow.---}
Firstly, we consider spin hydrodynamic generation for the two-dimensional Poiseuille flow in graphene. 
Spin diffusion equation in non-equilibirum steady state is given by\cite{SHD:Matsuo}
\begin{eqnarray}
\Big(\nabla^2 -\lambda^{-2}   \Big)\delta \mu_s = \hbar\lambda^{-2} \zeta\omega, \label{eq:SD}
\end{eqnarray}
where $\lambda=\sqrt{D_{\rm s} \tilde{\tau}_{\rm sf}} $ is the spin diffusion length.
The parallel flow between the graphene edges, $y=\pm y_0$, induces the velocity filed $\vv = ( v_0 (1-y^2/y_0{}^2),0)$, and then, the vorticity field becomes $\omega = \nabla \times \vv = 2v_0 y/y_0{}^2$.
In this case, the $z$-polarized spin current density reads\citep{SHD:Matsuo}
$j_{s,y}^z(y) 
= - \frac{\hbar^2 }{4e^2} \sigma_0 \zeta\frac{\rho}{\eta}E \Big[ 1-\frac{\cosh(y/\lambda)}{\cosh (y_0/\lambda)} \Big]$. 
Here we use the relation $v_0/y_0^2 = -\rho E/2\eta$. Total spin current for a strip with length $L$ and width $2y_0$ is given by
%
$J_{{\rm s},y}^z=-\frac{\hbar^2}{2e}\sigma_0\zeta\frac{\rho}{\eta}EL\left[y_0-\lambda\tanh\frac{y_0}{\lambda}\right]$.
%
The spin Hall angle is defined by $\theta_{\rm SH}= (2e/\hbar)J_{s,y}^z/J_{c,x}$ where the total charge current along the flow direction is given by
$J_{c,x} = -L\int_{-y_0}^{y_0}dy \rho v_x(y) = -\frac{4}{3}\rho v_0Ly_0
$. 
Thus, the spin Hall angle reads
\begin{eqnarray}
\theta_{\rm SH} = \frac{\hbar\zeta}{2m^*\nu}\left[1-\frac{\lambda}{y_0}\tanh\frac{y_0}{\lambda}\right],
\end{eqnarray}
where $m^*$ is the effective (cyclotron) electron mass and $\nu\equiv\eta/(nm^*)$ is the kinematic viscosity with the electron density $n$. 
Here, we use the relation of mobility, $\mu=-\bar{v}_x/E=-\sigma_0/(ne)$, with the averaged velocity of the electron fluid, $\bar{v}_x=(2/3)v_0$.

The kinematic viscosity relates to the Fermi velocity $v_{\rm F}$ and the electron collision mean free path $l_{\rm ee}$ as $\nu=v_{\rm F}l_{\rm ee}/4$~\cite{LandauKinetics}. For doped graphene in the Fermi liquid regime, the electron-electron collision mean free path is given by
%
$l_{\rm ee}^{-1}=\frac{\pi k_{\rm F}}{N}\left(\frac{T}{T_{\rm F}}\right)^2\ln\left(\frac{2T_{\rm F}}{T}\right)
$, 
where $N=4$ is the number of fermion flavors in graphene~\cite{Principi2016}.  For $y_0\gg\lambda$, the temperature dependence of the spin Hall angle,
\begin{align}
\theta_{\rm SH}=\frac{2\zeta}{k_{\rm F}l_{\rm ee}}=\frac{1}{4}\left(\frac{T}{T_{\rm F}}\right)\ln\left(\frac{2T_{\rm F}}{T}\right),
\end{align}
is shown in Fig.~\ref{fig:SHA}.
Note that the spin Hall angle is determined only by the normalized temperature $T/T_{\rm F}$ and is parameter free.

\begin{figure}[!hbtp]
	\begin{center}
   \includegraphics[scale=0.3]{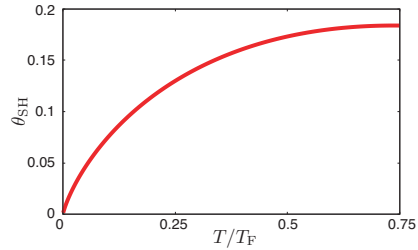}
	\caption{ Temperature dependence of the spin Hall angle for $y_0\gg\lambda$. 
    }\label{fig:SHA}
	\end{center}
\end{figure}

\begin{figure}[!hbtp]
	\begin{center}
   \includegraphics[scale=0.35]{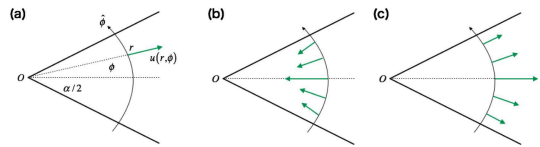}
	\caption{ Schematics of the Jeffery-Hamel flow. (a) Setup. (b) Convergent symmetrical flow is obtained for any Reynolds number.   (c) Divergent symmetrical flow is obtained for small Reynolds number. 
    }\label{fig:JH}
	\end{center}
\end{figure}

\paragraph{Spin current in Jeffery-Hamel flow.---}
Let us consider an analytical solution of the Navier-Stokes equation for a radial steady flow, known as the Jefferey-Hamel flow\cite{Jeffery1915,Hamel1917,LandauFluid}. 
The Jeffery-Hamel flow is the steady flow between two plane walls meeting at angle $\alpha$ (Fig. \ref{fig:JH}), and flows in/out from the line of intersection of the planes. 
We take  polar coordinates $(r, \phi)$ and assume the flow to be radial: 
$v_r=v(r,\phi), v_\phi =0$. Equation (\ref{eq:NS}) leads to
\begin{eqnarray}
&&v\frac{\partial v}{\partial r} = -\frac{1}{\tilde\rho}\frac{\del p}{\del r}+ \nu \Big( \frac{\del^2 v}{\del r^2} + \frac{1}{r^2} \frac{\del^2 v}{\del \phi^2} + \frac{1}{r}\frac{\del v}{\del r}  -\frac{v}{r^2}     \Big), \label{eq:23.5}\\
&&- \frac{1}{\tilde \rho r}\frac{\del p}{\del \phi} + \frac{2\nu}{r^2}\frac{\del v}{\del \phi} =0, \frac{\del (r v)}{\del r} = 0.
\end{eqnarray}
Thus, we obtain the analytical solution:
$v(r,\phi) = 6 \nu u(\phi)/r$,
and the vorticity becomes
$\omega = -\frac{1}{r} \frac{\del v(r,\phi)}{\del \theta} = -\frac{6\nu}{r^2} \frac{d u(\phi)}{d \phi}$.   
We introduce the mass $Q$ of fluid that passes in unit time through any cross-section $r=$constant:
$Q=\rho \int_{-\alpha/2}^{\alpha/2} d\phi \, r v = 6\nu \rho \int_{-\alpha/2}^{\alpha/2} d\phi  u(\phi)$.
The dimensionless parameter
$|Q|/\nu\rho$
plays a kind of the Reynolds number in this flow.

In Fig.~\ref{fig:JH}, the radial velocity is shown. 
\begin{figure}[!hbtp]
\begin{center}
   \includegraphics[scale=0.3]{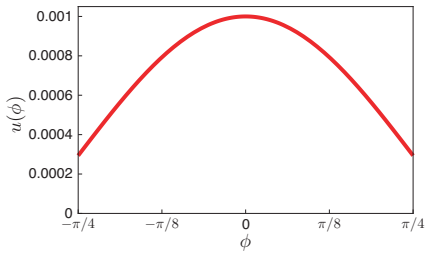}
	\caption{$\phi$ dependence of $u(\phi)$ in radial velocity $v(r,\phi)$ for the Jeffery-Hamel flow. The transverse axis means $\phi$ from $-\pi/4$ to $\pi/4$ and the longitudinal axis means $u(\phi)$. Here, we assume Re$=10^{-3}$.
    }\label{fig:JH_rad}
	\end{center}
\end{figure}
In the case of $Q<0$, we have convergent symmetrical flow for any Reynolds number. 
In contrast, when $Q>0$, we obtain divergent symmetrical flow only for small Reynolds number, namely, the flow becomes unstable and turbulent for a certain Reynolds number. 

Let us recall the spin diffusion eq. (\ref{eq:SD}). 
For simplicity, we assume $\lambda$ is much smaller than the system size. In this case, Eq. (\ref{eq:SD}) is easily solved as
$\delta \mu_s (r,\phi)\approx -\hbar \zeta \omega(r,\phi)$, 
and then, we obtain the spin current
\begin{eqnarray}
J_{s,r}= -\frac{\hbar\sigma_0}{4e^2} \frac{\del }{\del r}\delta \mu_s,
J_{s,\phi}= -\frac{\hbar\sigma_0}{4e^2}\frac{1}{r} \frac{\del }{\del \phi}\delta \mu_s. \label{eq:SC-r}
\end{eqnarray}
In a realistic situation, where the spin diffusion length is comparable to the system size, the spin diffusion equation has to be solved numerically. Nevertheless, an analytical solution exists for certain angles $\alpha$ and can be obtained by the method of images. For $\alpha =\pi/2$ the solution of the spin diffusion equation is given by
\begin{eqnarray}
\delta\mu_S(r,\phi)\!= \!\frac{6\zeta\hbar\nu}{\lambda^2}\!\!\int^{r_0}_0\!\!\frac{d\tilde{r}}{\tilde{r}}\!\!\int^{\alpha/2}_{-\alpha/2}\!\!\!\! d\tilde{\phi}G(r,\phi;\tilde{r},\tilde{\phi})\frac{du(\tilde{\phi})}{d\phi},
\label{eq:mu-Sol}
\end{eqnarray}
where $r_0$ is the length of the system and $G(r,\phi;\tilde{r},\tilde{\phi})$ is the Green's function for the spin diffusion equation given by 
$G(r,\phi;\tilde{r},\tilde{\phi})=\frac{1}{2\pi}\sum^2_{n=1}[K_0(R_n/\lambda)+K_0(\bar{R}_n/\lambda)]$
with $K_0(z)$ being the modified Bessel function of the second kind and $R_n$ and $\bar{R}_n$ being $R_n=\sqrt{r^2-2r\tilde{r}\cos(\phi-n\pi+\tilde{\phi})+\tilde{r}^2}$ and $\bar{R}_n=\sqrt{r^2+2r\tilde{r}\cos(\phi-n\pi-\tilde{\phi})+\tilde{r}^2}$, respectively. Here, the Neumann boundary condition is used.

\begin{figure}[!hbtp]
	\begin{center}
   \includegraphics[scale=0.4]{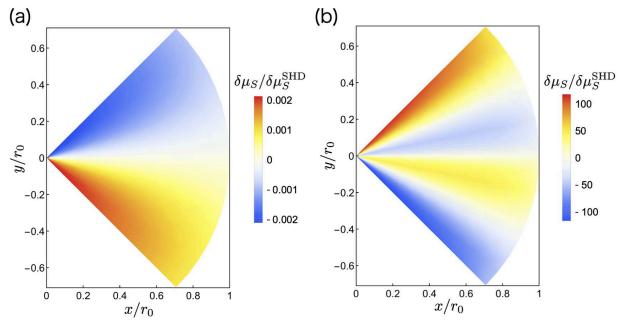}
	\caption{ Numerical solutions of the spin diffusion equation for the Jeffery-Hamel flow for (a) Re$=10^{-3}$ and (b) Re$=100$. Color plot of the spin accumulation $\delta \mu_s$ as the function of $\phi$ and $r$ in the case of $\alpha = \pi/2$. The data are normalized by the factor $\delta\mu^{\textrm{SHD}}_S:=6\zeta\hbar\nu/\lambda^2$, estimated as $0.1$~meV. The blue and red regions describe the negative and positive spin accumulation, respectively.  
    }\label{fig:JH_dmus_R0001}
	\end{center}
\end{figure}

The spin accumulation caused by the Jeffery-Hamel flow is shown in Fig.~\ref{fig:JH_dmus_R0001}. Here, we show the numerical result of Eq.~(\ref{eq:mu-Sol}) normalized the factor $6\zeta\hbar\nu/\lambda^2$ for two dimensional space. The function $u(\phi)$ in the radial velocity is shown in Fig.~\ref{fig:JH_rad}.  Figure \ref{fig:JH_dmus_R0001} shows that the positive tangential component of the spin current $J_{s,\phi}$, defined in Eq.~(\ref{eq:SC-r}) flows while the radial component of the spin current $J_{s.r}$, defined in Eq.~(\ref{eq:SC-r}) is relatively small since the system size $r_0$ is comparable to the spin relaxation length $\lambda$. 

\paragraph{Discussion and conclusion.---}
Let us first compare the efficiency of the charge-to-spin conversion in graphene emerging as a result of the spin-vorticity coupling with that stemming from the spin-orbit coupling. Fig.~\ref{fig:SHA} shows that the spin Hall angle $\theta_{\rm SH}$ exceeds 0.1 over a wide range of temperatures, a value that is comparable to that found in CVD graphene due to the scattering from residual copper adatom clusters \cite{SHE:Graphene-particles}, and significantly larger than that reported in Ref. \cite{SLG:SHA2014}, where the spin Hall angle due to induced spin-orbit coupling at room temperature is estimated to be $6.1 \times 10^{-7}$, and 
Next, in graphene $\nu=0.1$~m$^2$/s~\cite{Bandurin2016} and $\lambda=2$~$\mu$m~\cite{Tombros07} at $T=0.2\,T_{\rm F}$, and therefore the factor $6\zeta\hbar\nu/\lambda^2$ in the Jeffery-Hamel flow is of the order of $0.1$~meV (or $0.83$~T). The maximum value in Fig.~\ref{fig:JH_dmus_R0001} is of the order of $2.6\times10^{-4}$ 
, so that the spin accumulation reaches $3\times10^{-2}$~$\mu$eV (or $1$~Oe) at the sample boundaries. This value is comparable to that observed in the spin Peltier effect~\cite{Ohnuma17} and is within the sensitivity of the modern magnetometry techniques.

The spin accumulation, in the case of Jeffery-Hamel flow, can be conveniently controlled by the set of the following parameters. First of all, depending on the sign of $Q$, the flow can be either convergent or divergent. In the latter case, the flow may become unstable for high values of the Reynolds number which may cause the onset of turbulence \cite{turbulence} in the electron fluid and a peculiar distribution of spin accumulation which we plot in Fig. \ref{fig:JH_dmus_R0001}. Furthermore, by changing charge carrier density and temperature, the viscosity of the electron fluid can be varied, which will directly affect the absolute value of the spin accumulation. Last but not least, an angle between the sample edges is another parameter which is responsible for the magnitude of the spin accumulation. In particular, $\alpha \to \pi$ corresponds to the geometry studied in Ref. \cite{Bandurin2016} and therefore our results indicate the possibility to observe the hydrodynamic spin Hall effect in conventional graphene Hall bars. 

In conclusion, we studied the generation of spin current as a result of viscous hydrodynamic flow of electrons in graphene. By combining the Navier-Stoke equations and the spin diffusion equation in the presence of spin-vorticity coupling, spin current generation has been predicted to emerge in the case of parallel laminar flow and Jeffery-Hamel flow. In the parallel flow case, we defined the spin Hall angle and estimated it to exceed 0.1 over the wide range of temperatures, a value close to that for materials with large spin-orbit coupling. In the Jeffery-Hamel flow configuration, we studied the dependence of spin current generation and spin accumulation on various parameters such as carrier density and sample geometry. We also provided quantitative estimates for the spin Hall angle and spin accumulation and found that the obtained values are within the sensitivity of some experimental techniques such as scanning NV-magnetrometry~\cite{NVcent} and therefore can be probed directly. 
Another possible experimental demonstration is the two-dimensional sensing technique\cite{Riverside}. D. Mayers et al. proposed a scanning photovoltage microscopy technique\cite{Riverside}, which enables the spatial mapping of electron flow in spintronic devices.  By applying theses techniques, we may measure the conversion from spin current to hydrodynamic electron flow.
Our result reveals a new functionality of graphene as a promising spin current source. 

\paragraph{Acknowledgements.---}
M.M. is financially supported by the Priority Program of Chinese Academy of Sciences, Grant No. XDB28000000, and KAKENHI (No. 20H01863) from MEXT, Japan.
Y.T. is financially supported by ERATO, JST. S.M. is financially supported by KAKENHI (No. 26103005, No. JP16H04023, and No. JP26247063) from MEXT, Japan. Theoretical work of D.B. was supported by the Russian Science Foundation: Grant 18-72-00234 (hydrodynamic theory).



\begin{thebibliography}{10}
\bibitem{Zutic2004}I. Zutic, J. Fabian, S. Das Sarma, 
Rev. Mod. Phys. {\bf 76}, 323 (2004).
\bibitem{MaekawaEd2012}S. Maekawa, S. Valenzuela, E. Saitoh, and T. Kimura, ed., {\it Spin Current} (Oxford University Press, Oxford, 2012).

\bibitem{Jedema2001}F.J. Jedema, A.T. Filip, B.J. van Wees, 
Nature (London) {\bf 410}, 345 (2001).

\bibitem{Mizukami2002}S. Mizukami, Y. Ando, T. Miyazaki, 
Phys. Rev. B {\bf 66}, 104413 (2002).

\bibitem{Tserkovnyak2005}Y. Tserkovnyak, A. Brataas, Gerrit E. W. Bauer, Bertrand I. Halperin, 
Rev. Mod. Phys. {\bf 77}, 1375 (2005).


\bibitem{SSE1}K. Uchida, S. Takahashi, K. Harii, J. Ieda, W. Koshibae, K. Ando, S. Maekawa, E. Saitoh, 
Nature {\bf 455}, 778 (2008). 


\bibitem{SSE1a}K. Uchida, J. Xiao, H. Adachi, J. Ohe, S. Takahashi, J. Ieda, T. Ota, Y. Kajiwara, H. Umezawa, H. Kawai, G. E. W. Bauer, S. Maekawa, E. Saitoh, 
Nat. Mater. {\bf 9}, 894 (2010).

\bibitem{SSE2} C. M. Jaworski, J. Yang, S. Mack, D. D. Awschalom, J. P. Heremans, R. C. Myers, 
Nat. Mater. {\bf 9}, 898 (2010).

\bibitem{ASP}K. Uchida, H. Adachi, T. An, T. Ota, M. Toda, B. Hillebrands, S. Maekawa, E. Saitoh, 
Nat. Mater. {\bf 10}, 737 (2011). 

\bibitem{ASP2}K. Uchida, T. An, K. Kajiwara, M. Toda, E. Saitoh, 
Appl. Phys. Lett. {\bf 99}, 212501 (2011). 

\bibitem{ASP3}K. Uchida, H. Adachi, T. An, H. Nakayama, M. Toda, B. Hillebrands, S. Maekawa, E. Saitoh, 
J. Appl. Phys. {\bf 111}, 053903 (2012).

\bibitem{Valenzuela2006}S. O. Valenzuela, M. Tinkham, 
Nature (London) {\bf 442}, 176 (2006).
\bibitem{Saitoh2006}E. Saitoh, M. Ueda, H. Miyajima, G. Tatara, 
Appl. Phys. Lett. {\bf 88}, 182509 (2006).
\bibitem{NLSV}T. Kimura, Y. Otani, T. Sato, S. Takahashi, S. Maekawa, 
Phys. Rev. Lett. {\bf 98}, 156601 (2007).

\bibitem{SAW-SC:Matsuo}M. Matsuo, J. Ieda, K. Harii, E. Saitoh, S. Maekawa,
Phys. Rev. B {\bf 87}, 180402(R) (2013). 

\bibitem{Takahashi2016}R. Takahashi, M. Matsuo, M. Ono, K. Harii, H. Chudo, S. Okayasu, J. Ieda, S. Maekawa, E. Saitoh, 
Nat. Phys. {\bf 12}, 52 (2016). 

\bibitem{Kobayashi2017}D. Kobayashi, Y. Yoshikawa, M. Matsuo, R. Iguchi, S. Maekawa, E. Saitoh, Y. Nozaki, 
Phys. Rev. Lett. {\bf 119}, 077202 (2017).

\bibitem{Okano2019}G. Okano, M. Matsuo, Y. Ohnuma, S. Maekawa, and Y. Nozaki,
Phys. Rev. Lett. {\bf 122}, 217701 (2019).


\bibitem{Hehl1976}F.W. Hehl, P. von der Heyde, G.D. Kerlick, J.M. Nester,
Rev. Mod. Phys. {\bf 48}, 393, (1976). 



\bibitem{Kawakami2014}W. Han, R. K. Kawakami, M. Gmitra, J.Fabian, 
Nature Nanotechnology {\bf 9}, 794-807 (2014).

\bibitem{Drogler}M. Drögeler, F. Volmer, M. Wolter, B. Terrés, K. Watanabe, T. Taniguchi, G. Güntherodt, C. Stampfer, B. Beschoten, 
Nano Lett. 14(11), 6050-6055 (2014).

\bibitem{Ingla-Aynes2015}J. Ingla-Aynes, M. H. D. Guimares, R. J. Meijerink, P. J. Zomer, B. J. Van Wees, 
Phys. Rev. B 92, 20, (2015).

\bibitem{Yan2016}J. W. Yan, O. Txoperena, R. Llopis, H. Dery, L. E. Hueso, F. Casanova, 
Nature Communications 7, 13372 (2016).

\bibitem{Tombros07} N. Tombros, C. Jozsa, M. Poponciuc, H. T. Jonkman, B. J. van Wees, 
Nature {\bf 448}, 571 (2007).

\bibitem{Balakrishnan}J. Balakrishnan, G. Kok Wai Koon, M. Jaiswal, A. H. Castro Neto, B. Özyilmaz, 
Nature Physics {\bf 9}, 284–287 (2013).





\bibitem{Crossno2016}J. Crossno, J. K. Shi, K. Wang, X. Liu, A. Harzheim, A. Lucas, S. Sachdev, P. Kim, T. Taniguchi, K. Watanabe, T. A. Ohki, K. C. Fong,
Science {\bf 351}, 1058 (2016).

\bibitem{Moll2016} J. W. Moll, P. Kushwaha, N. Nandi, B. Schmidt, A. P. Mackenzie, 
Science 351, 1061 (2016).

\bibitem{Bandurin2016}D. A. Bandurin, I. Torre, R. Krishna Kumar, M. Ben Shalom,
A. Tomadin, A. Principi, G. H. Auton, E. Khestanova, K. S. Novoselov, I. V. Grigorieva, L. A. Ponomarenko, A. K. Geim, M. Polini, 
Science {\bf 351}, 1055 (2016).


\bibitem{gurzhi63}
R. N. Gurzhi,
{\it Usp. Fiz. Nauk} {\bf 94}, 689 [Engl. transl.: {\it Sov. Phys. Usp.} {\bf 11}, 255 (1968)].
\bibitem{Lucas2018}A. Lucas, K. C. Fong, 
Journal of Physics: Condensed Matter {\bf 30}, 5 (2018).

\bibitem{Levitov2016}L. Levitov, G. Falkovich, 
Nat. Phys. {\bf 12}, 672–676 (2016).

\bibitem{Bandurin2018}D. A. Bandurin, A.V. Shytov, L.S. Levitov, R. Kirshna Kumar, A. I. Berdyugin, M. Ben Shalom, I. V. Grigorieva, A. K. Geim, G. Falkovich, 
Nature Communications {\bf 9}, 4533 (2018).

\bibitem{Torre2015} I. Torre, A. Tomadin, A. K. Geim, and  M. Polini, 
Phys. Rev. B {\bf 92}, 165433 (2015).

\bibitem{Vozmediano}M.A.H. Vozmediano, M.I. Katsnelson, and F. Guinea, 
Phys. Rep. 496 (2010) 109–148.

\bibitem{Guo}  H. Guo, E. Ilseven, G. Falkovich, L. Levitov, 
Proc. Natl. Acad. Sci. U.S.A 114(12) 3068-3073 (2017)

\bibitem{Kumar2017} R. K. Kumar, D. A. Bandurin, F. M. D. Pellegrino, Y. Cao, A. Principi, H. Guo, G. H. Auton, M. Ben Shalom, L. A. Ponomarenko, G. Falkovich, K. Watanabe, T. Taniguchi, I. V. Grigorieva, L. S. Levitov, M. Polini, A. K. Geim, 
Nat. Phys. {\bf 13}, 1182 (2017).
 

\bibitem{Gallagher2019}P. Gallagher, C.-S. Yang, T. Lyu, F. Tian, R. Kou, H. Zhang, K. Watanabe, T. Taniguchi, F. Wang, 
Science {\bf 364}, 6436, 158-162 (2019)


\bibitem{Berdyugin2018}
A. I. Berdyugin, S. G. Xu, F. M. D. Pellegrino, R. Krishna Kumar, A. Principi, I. Torre, M. Ben Shalom, T. Taniguchi, K. Watanabe, I. V. Grigorieva, M. Polini, A. K. Geim, D. A. Bandurin, 
Science {\bf364}, 6436, 162-165 (2019)


\bibitem{HydroGraphene}M. Mendoza, H. J. Hermann, S. Succi, 
Scientific Reports {\bf 3}, 1052 (2013).

\bibitem{SHD:Matsuo}M. Matsuo, Y. Ohnuma, S. Maekawa, 
Phys. Rev. B {\bf 96}, 020401(R) (2017). 

\bibitem{RenormSRC} M. Matsuo, J. Ieda, S. Maekawa, 
Phys. Rev. B {\bf 87}, 115301 (2013).

\bibitem{Doornenbal2019}
R. J. Doornenbal, M. Polini, R. A. Duine, 
J. Phys. Mater. {\bf 2}, 015006 (2019).


\bibitem{LandauKinetics}E. M. Lifshitz, L.P. Pitaevskii, Physical Kinetics (Pergamon Press, 1987).




\bibitem{Principi2016} A. Principi, G. Vignale, C. Matteo, M. Polini,
Phys. Rev. B {\bf 93}, 125410 (2016).

\bibitem{Jeffery1915}
Jeffery, G. B. "L. The two-dimensional steady motion of a viscous fluid." The London, Edinburgh, and Dublin Philosophical Magazine and Journal of Science 29.172 (1915): 455–465.
\bibitem{Hamel1917}
Hamel, Georg. "Spiralförmige Bewegungen zäher Flüssigkeiten." Jahresbericht der Deutschen Mathematiker-Vereinigung 25 (1917): 34–60.

\bibitem{LandauFluid}L. D. Landau, E. M. Lifshitz, {\it Fluid Mechanics}  (Pergamon, Oxford, 1987).




\bibitem{SHE:Graphene-particles}
%
J. Balakrishnan, G. Kok Wai Koon, A. Avsar, Y. Ho, J. Hak Lee, M. Jaiswal, S.-J. Baeck, J.-H. Ahn, A. Ferreira, M. A. Cazalilla, A. H. Castro Neto, B. Özyilmaz, 
Nature Communications {\bf 5}, 4748 (2014)

\bibitem{SLG:SHA2014}
R. Ohshima, A. Sakai, Y. Ando, T. Shinjo, K. Kawahara, H. Ago, M. Shiraishi, 
Appl. Phys. Lett. {\bf 105}, 162410 (2014).


\bibitem{Ohnuma17}Y. Ohnuma, M. Matsuo, S. Maekawa, 
Phys. Rev. B {\bf 96}, 134412 (2017).

\bibitem{turbulence}
A. Gabbana, M. Polini, S. Succi, R. Tripiccione, and F. M. D. Pellegrino, 
Phys. Rev. Lett. {\bf 121}, 236602 (2018).

\bibitem{NVcent}J. R. Maze, P. L. Stanwix, J. S. Hodges, S. Hong, J. M. Taylor, P. Cappellaro, L. Jiang, M. V. Gurudev Dutt, E. Togan, A. S. Zibrov, A. Yacoby, R. L. Walsworth, M. D. Lukin, 
Nature {\bf455}, 644–647 (2008).

\bibitem{Riverside}
D. Mayes, M. Grossnickle, M. Lohmann, M. Aldosary, J. Li, V. Aji, J. Shi, J. C.W. Song, and N. M. Gabor,
arXiv:2002.07902.



\end{thebibliography}
\end{document}